%% file: main.tex
\documentclass[10pt, conference]{IEEEtran}
\IEEEoverridecommandlockouts
\usepackage{balance}
\usepackage{ulem}
\usepackage{cite}
\usepackage{amsmath,amssymb,amsfonts}
\usepackage{algorithmic}
\usepackage[ruled, linesnumbered]{algorithm2e}
\usepackage{graphicx}
\usepackage{textcomp}
\usepackage{xcolor}
\usepackage{color}
\usepackage{url}
\usepackage{enumitem}
\usepackage{multirow}
\usepackage{verbatim}
\usepackage{soul}
\usepackage{listings}
\usepackage{caption}
\usepackage{subcaption}
\usepackage{bbding}
\usepackage{colortbl}
\usepackage{pgfplots}
\usepackage{titlesec}
\usepackage{microtype}
\usepackage{pifont}
\usepackage{flushend}
\usepackage{threeparttable}
\usepackage{hyperref}
\usepackage{booktabs}
\usepackage{tcolorbox}
\usepackage{changepage}
\usepackage{inconsolata}
\usepackage{lipsum}
\usepackage{pifont}
\usepackage{orcidlink}

\soulregister{\ref}7 

\definecolor{mypurple}{HTML}{0645AD}
\definecolor{mymask}{HTML}{ffd700}
\definecolor{mygreen}{HTML}{6E8B3D}
\definecolor{mycommentcolor}{HTML}{0671b9}
\newcommand{\greenbf}[1]{{\color[HTML]{458B00}\textbf{\texttt{#1}}}}

\newcommand{\mycommentstyle}[1]{\color[HTML]{0671b9}{\small #1}}
\SetKwComment{Comment}{\mycommentstyle{\# }}{}
\newcommand{\tool}{\textsc{ClozeMaster}}

\newcommand{\mypara}[1]{\smallskip\noindent\textit{\textbf{#1} }}

\newcommand{\reported}{37}
\newcommand{\confirmed}{27}
\newcommand{\fixed}{10}
\newcommand{\duplicate}{4}
\newcommand{\wontfix}{4}

\newcommand{\rhreported}{15}
\newcommand{\rhconfirmed}{15}
\newcommand{\rhfixed}{3}
\newcommand{\rhduplicate}{0}
\newcommand{\rhwontfix}{0}

\newcommand{\rireported}{17}
\newcommand{\riconfirmed}{9}
\newcommand{\rifixed}{5}
\newcommand{\riduplicate}{4}
\newcommand{\riwontfix}{4}

\newcommand{\mireported}{5}
\newcommand{\miconfirmed}{3}
\newcommand{\mifixed}{2}
\newcommand{\miduplicate}{0}
\newcommand{\miwontfix}{0}

\newcommand{\mhreported}{0}
\newcommand{\mhconfirmed}{0}
\newcommand{\mhfixed}{0}
\newcommand{\mhduplicate}{0}
\newcommand{\mhwontfix}{0}

\newcommand{\keepnotes}{true} 

\ifthenelse{\equal{\keepnotes}{true}}{
  \newcommand{\mytodogreen}[1]{\textcolor{mygreen}{\ding{46}~{\sf}~#1}}
\newcommand{\yyb}[1]{\textcolor{red}{[yyb: #1]}}
\newcommand{\sml}[1]{\textcolor{blue}{[sml: #1]}}
}{
  \newcommand{\mytodogreen}[1]{}
  \newcommand{\yyb}[1]{}
  \newcommand{\sml}[1]{}
}

\newtcbox{\mybox}[1][darkgray]
  {on line, arc = 0pt, outer arc = 0pt,
    colback = #1!10!white, colframe = #1!50!black,
    boxsep = 0pt, left = 1pt, right = 1pt, top = 2pt, bottom = 2pt,
    boxrule = 0pt, bottomrule = 0pt, toprule = 0pt}

\makeatletter
\def\url@leostyle{%
  \@ifundefined{selectfont}{\def\UrlFont{\sf}}{\def\UrlFont{\small\ttfamily}}}
\makeatother
\def\BibTeX{{\rm B\kern-.05em{\sc i\kern-.025em b}\kern-.08em
    T\kern-.1667em\lower.7ex\hbox{E}\kern-.125emX}}
\begin{document}

\title{ClozeMaster: Fuzzing Rust Compiler by Harnessing LLMs for Infilling Masked Real Programs}


\author{
\IEEEauthorblockN{
Hongyan Gao,
Yibiao Yang\textsuperscript{†},
Maolin Sun,
Jiangchang Wu,
Yuming Zhou,
Baowen Xu
}
\IEEEauthorblockA{
\textit{State Key Laboratory for Novel Software Technology, Nanjing University, Nanjing, China}\\
\{hongyangao2023, merlin, jiangchangwu\}@smail.nju.edu.cn\\
\{yangyibiao, zhouyuming, bwxu\}@nju.edu.cn
}
}

\maketitle

\renewcommand{\thefootnote}{\fnsymbol{footnote}}
\setcounter{footnote}{2}   
\footnotetext{Corresponding author: Yibiao Yang (yangyibiao@nju.edu.cn)}

\begin{abstract}
\label{sec:abstract}
\input{content/abstract.tex}
\end{abstract}


\section{Introduction}
\label{sec:intro}
\input{content/introduction.tex}


\section{Background \& Motivation}
\label{sec:back}
\input{content/back_movti.tex}

\section{Approach}
\label{sec:approach}
\input{content/approach.tex}

\section{Evaluation}
\label{sec:evaluation}
\input{content/evaluation.tex}

\section{Discussion}
\label{sec:discussion}
\input{content/discussion.tex}

\section{Related Work}
\label{sec:related}
\input{content/related.tex}

\section{Conclusion}
\label{sec:conclusion}
\input{content/conclusion.tex}

\section*{Acknowledgments}

We are grateful to the anonymous reviewers for their insightful suggestions on earlier versions of this paper. We are also indebted to the Rust compiler developers for inspecting and fixing our reported bugs. 
Yibiao Yang is the corresponding author. 
This work is partially supported by the National Natural Science Foundation of China (62072194, 624B2067, 62172205, and 62272214), the Jiangsu Natural Science Foundation (Grant BK20231402), the Collaborative Innovation Center of Novel Software Technology and Industrialization, and the Fundamental Research Funds for the Central Universities (14380121).

\balance
\bibliographystyle{IEEEtran}
\normalem
\bibliography{full}

\end{document}

%% file: content/abstract.tex
Ensuring the reliability of the Rust compiler is of paramount importance, given increasing adoption of Rust for critical systems development, due to its emphasis on memory and thread safety. 
However, generating valid test programs for the Rust compiler poses significant challenges, given Rust's complex syntax and strict requirements. With the growing popularity of large language models (LLMs), much research in software testing has explored using LLMs to generate test cases. Still, directly using LLMs to generate Rust programs often results in a large number of invalid test cases. Existing studies have indicated that test cases triggering historical compiler bugs can assist in software testing. Our investigation into Rust compiler bug issues supports this observation. Inspired by existing work and our empirical research, 
we introduce a bracket-based masking and filling strategy called \texttt{clozeMask}. 
The \texttt{clozeMask} strategy involves extracting test code from historical issue reports, identifying and masking code snippets with specific structures, and using an LLM to fill in the masked portions for synthesizing new test programs. This approach harnesses the generative capabilities of LLMs while retaining the ability to trigger Rust compiler bugs.
It enables comprehensive testing of the compiler's behavior, particularly exploring edge cases.
We implemented our approach as a prototype \tool. \tool~has identified \confirmed~confirmed bugs for \textit{rustc} and \textit{mrustc}, of which \fixed~have been fixed by developers. 
Furthermore, our experimental results indicate that \tool~outperforms existing fuzzers in terms of code coverage and effectiveness.

%% file: content/introduction.tex
Rust, a modern, safe, and concurrent systems-level programming language~\cite{HowDoProgrammersUseUnsafeRust,10.1145/3586037,10.1145/3591335.3591340}, has garnered significant attention in both the software development and national-level cybersecurity fields in recent years~\cite{SyRust,Rustsmith,10.1145/3622821,10.1145/3622841,10.1109/ASE51524.2021.9678813,lwm,microsoft,huggingface,app}. According to a GitHub survey, Rust is the second fastest-growing programming language~\cite{top.language.2022}. The White House administration advocates for developers to ``preferentially use memory-safe languages such as Rust" in its latest report~\cite{usa.report} in 2024, indicating the growing importance and potential of Rust in the software development and cybersecurity domains.

The adoption of Rust has increased, making bugs in the Rust compiler a significant concern for client users. Unnoticed errors during compilation can lead to undefined behavior or crashes at runtime, posing a serious risk for organizations relying on the Rust compiler for system development and critical tasks. Hence, systematically testing the Rust compiler is crucial~\cite{Rustsmith}.

Prior work on compiler testing can be classified into two categories: \ding{182}generation-based fuzzers and \ding{183}mutation-based fuzzers~\cite{SynFuzz}. Generation-based fuzzers construct test cases using random syntax rules or templates~\cite{Csmith,Yarpgen,CLSmith}, while mutation-based fuzzers generate additional test cases by creating equivalent programs or applying mutation templates and heuristics~\cite{EMI,SLEMI,10.1145/2813885.2737986,10.1145/3527317,spe}. These methods have demonstrated effectiveness in testing compilers for mature languages like C/C++. However, due to the complex and diverse language features of Rust, developing a comprehensive generative fuzzer that covers its entire semantic landscape is challenging and time-consuming~\cite{Rustsmith,RustHardSurvey,Coblenz2023}. Furthermore, the evolving nature of Rust and the limited maturity of its analysis tools make implementing mutation-based fuzzers difficult~\cite{ARDITO2020100635}.

Recently, the development of large language models (LLMs) has brought new advances to the field of software testing~\cite{Scinf2024-deep-learning-survey,whitefox,FuzzGpT,TitanFuzz,KernelGPT,UniverseFuzz}. Titanfuzz proposes the direct generation of code snippets using LLM, synthesizing Python test cases for deep learning libraries by masking function bodies or parameters based on output results~\cite{TitanFuzz}. FuzzGPT fine-tunes LLM by collecting code segments that historically triggered bugs in deep learning libraries, generating Python test cases using textual prompts~\cite{FuzzGpT}. Fuzz4all and KernelGPT employ carefully designed prompts to guide LLM in generating test code for specific software or kernel programs~\cite{UniverseFuzz,KernelGPT}. Despite notable achievements, these methods still face limitations in generating test programs for the Rust compiler.
For instance, Titanfuzz and Fuzz4all rely on LLM for zero-shot learning or generating test programs based solely on textual prompts. However, due to Rust's status as an emerging programming language, existing open-source LLMs mostly generate meaningless or invalid Rust programs without considering program context. Although FuzzGPT addresses this by fine-tuning LLMs with code segments that historically triggered bugs, directly generating effective Rust compiler test programs solely based on the fine-tuned model remains limited due to the small scale of Rust programs~\cite{incoder}.

Inspired by prior research~\cite{FuzzGpT,HistIssue,10.1109/ICSE48619.2023.00018}, we observed that test code historically associated with software bugs exhibits potential characteristics for triggering bugs in software systems. Motivated by this, we analyzed code snippets that have historically triggered bugs in the Rust compiler. Considering the limitations of existing test case generation approaches using LLMs, we present our effective approach for the Rust compiler.

\mypara{Approach.}We introduce \tool, an innovative fuzzing framework specifically designed for testing the Rust compiler. 
\tool~harnesses the power of LLMs and historical bug-triggering test inputs to address the unique challenges posed by the rigorous syntax of the Rust language. 
The core strategy adopted by \tool~is a \texttt{clozeMask} approach. 
It involves masking specific code segments within parentheses in historical bug-triggering code snippets.
Subsequently, an LLM is utilized to infill the masked code, completing the snippets. 
This process enables \tool~to generate new and potential bug-triggering test programs for the Rust compiler, leveraging the LLM's capabilities for code completion. 
Our key insight is that historically bug-triggering code snippets encapsulate valuable domain-specific knowledge, which can effectively explore edge cases in the Rust compiler. 
Similar strategies have proven successful in testing other software systems, such as C compiler~\cite{HistIssue} and SMT solvers~\cite{10.1109/ICSE48619.2023.00018}. 
To maximize the utilization of the rich information present in historical bug-triggering inputs, \tool~adoptes two-fold approach. 
Firstly, existing bug-triggering code snippets are collected, and specific code within parentheses is masked, serving as a foundation for generating new test code. 
Secondly, these bug-triggering code snippets are employed to fine-tune the LLM, enabling it to learn the syntax and semantic knowledge of code that can trigger Rust compiler bugs.
This approach enhances the effectiveness of \tool~in generating relevant and potentially bug-triggering test cases. 
To evaluate the effectiveness of \tool, we applied it to two practical Rust compilers actively used by developers: \textit{rustc}, the official Rust compiler, and \textit{mrustc}, an alternative Rust compiler developed by an individual developer. 
By employing \tool, we successfully identified have reported \reported~bugs, out of which \confirmed~are confirmed by developers. 
Notably, \fixed~of the \confirmed~bugs have been fixed. 
Furthermore, compared to existing Rust compiler testing tools, \tool~achieved higher code coverage and enhanced bug detection capabilities. 

\mypara{Contributions.} We make the following major contributions:
\begin{itemize}[leftmargin=*]
    \item We analyze the bug reports in the Rust compiler and introduce a simple yet effective fuzzing approach named \texttt{clozeMask}, which involves masking the code segment within paired parentheses in the historical bug-triggering code snippets.
    Our proposed \texttt{clozeMask} strategy harnesses LLM's capabilities in code completion and maintains the code snippet's bug-triggering potential. 
    \item We have implemented our proposed strategy in the form of a prototype called \tool. 
    \tool, using LLM as a cloze master, can serve as a general and practical fuzzer for the Rust compiler.
    \item
    We conducted an extensive evaluation for \tool~by applying it to two Rust compilers, namely \textit{rustc} and \textit{mrustc}. 
    \tool~successfully identified and reported \reported~bugs within these compilers, of which \confirmed~are confirmed and \fixed~of those confirmed bugs are fixed by developers. 
    This underscores the effectiveness of \tool~in exposing bugs for Rust compiler. 
 \end{itemize}

\mypara{Artifacts.}
The implementation of \tool, as well as the list of bugs we have identified, have been publicly available at: 
\href{https://github.com/clozeMasterPro/clozeMaster}{\textcolor{mypurple}{\textbf{https://github.com/clozeMasterPro/clozeMaster}}}

\mypara{Paper Organization.} 
The rest of this paper is structured as follows. 
Section \ref{sec:back} presents an overview of Rust language, large language models, empirical analysis of Rust compiler bug reports, and the resulting motivation.
Section \ref{sec:approach} formalizes our approach and describes the implementation of~\tool. 
Next, we elaborate on our extensive evaluation in detail in Section \ref{sec:evaluation}.
In Section \ref{sec:discussion}, we formulate more discussions on our results.
Finally, we survey related work in Section \ref{sec:related} and the conclusion is in Section \ref{sec:conclusion}.

%% file: content/back_movti.tex
 This section first presents a brief introduction to the Rust programming language and Large Pre-trained Language Models. Then, we motivate and illustrate our technique using empirical study and real examples.
\subsection{Background}
\subsubsection{Rust Language}

Rust, a systems programming language developed by Mozilla Research, was officially released in 2015 with the goal of providing solutions for high performance, memory safety, and concurrency~\cite{10.1145/3611643.3616303,10.1145/3611643.3613878}. Compared to other programming languages, Rust exhibits higher syntactic complexity, primarily due to its ownership model and borrowing rules. 
Rust's ownership system handles memory with strict borrowing checks to ensure safety and prevent errors like null pointers and data races. It also features lifetimes for managing borrowing, requiring annotations with \texttt{<'a>} for safe access.
While the syntactic features introduced by the Rust language significantly improve the safety of programs, their complex coding rules have also led Rust to be perceived as the ``too intimidating, too hard to learn, or too complicated" in a survey among developers in the Rust community~\cite{RustHardSurvey}. Furthermore, the stringent analysis enforced by the Rust compiler and the language's intricate syntax constraints reduce the likelihood of user-written code being accepted by the compiler~\cite{Coblenz2023}.

Consequently, the steep learning curve of the Rust language and the rigorous syntax checks enforced by the Rust compiler make it highly challenging to generate Rust test programs through traditional methods reliant on manually constructed templates or rules~\cite{10.1145/3510003.3510164,Coblenz2023,Rustsmith,CLPRust}.
\subsubsection{Large Pre-trained Language Models}
Large Language Models (LLMs) have demonstrated impressive performance across multiple domains within Natural Language Processing (NLP), trained on massive internet-extracted text data~\cite{10.5555/3495724.3495883,10.5555/3455716.3455856,Chowdhery2022PaLMSL}. These models employ deep learning techniques, typically based on the Transformer architecture. The Transformer effectively captures contextual relationships and generates coherent outputs. For example, in code generation or code completion tasks, LLMs undergo supervised learning, and training on extensive code data to grasp syntax, structure, and common patterns~\cite{starcoder,incoder,Wang2023CodeT5OC}. By maximizing accuracy in predicting the next code token, the model gradually improves understanding and generation of code semantics. Recently, researchers have leveraged pre-trained LLMs from open-source libraries, employing carefully designed prompt strategies or fine-tuning techniques for specific tasks like domain adaptive code generation~\cite{Tang2023DomainAC}.

Although LLMs perform well on many NLP tasks, they struggle with downstream tasks of generating compiler test programs for emerging languages like Rust due to the scarcity of training data. Analysis of models such as StarCoder and Incoder shows that Rust constitutes less than 1/20th of their pre-training corpora~\cite{starcoder,incoder}, leading to a long-tail problem~\cite{10.1109/TPAMI.2023.3268118}. Directly generating Rust programs with general-purpose ChatGPT4 is also challenging as it rarely triggers compiler bugs~\cite{FuzzGpT}. To address this issue, we have designed the \texttt{clozeMask} strategy to more effectively harness the power of LLMs for Rust-specific test code generation tasks.

\subsection{Preliminary Study}
We conduct a preliminary study on bug issues in the \textit{rustc} compiler to understand typical characteristics of test cases historically triggering Rust compiler bugs. From this data analysis, we make two important observations.

\subsubsection{Statistics of rustc Issues}
\label{subsec:preStudy}

\begin{figure}
    \centering
    \includegraphics[width=0.4\textwidth]{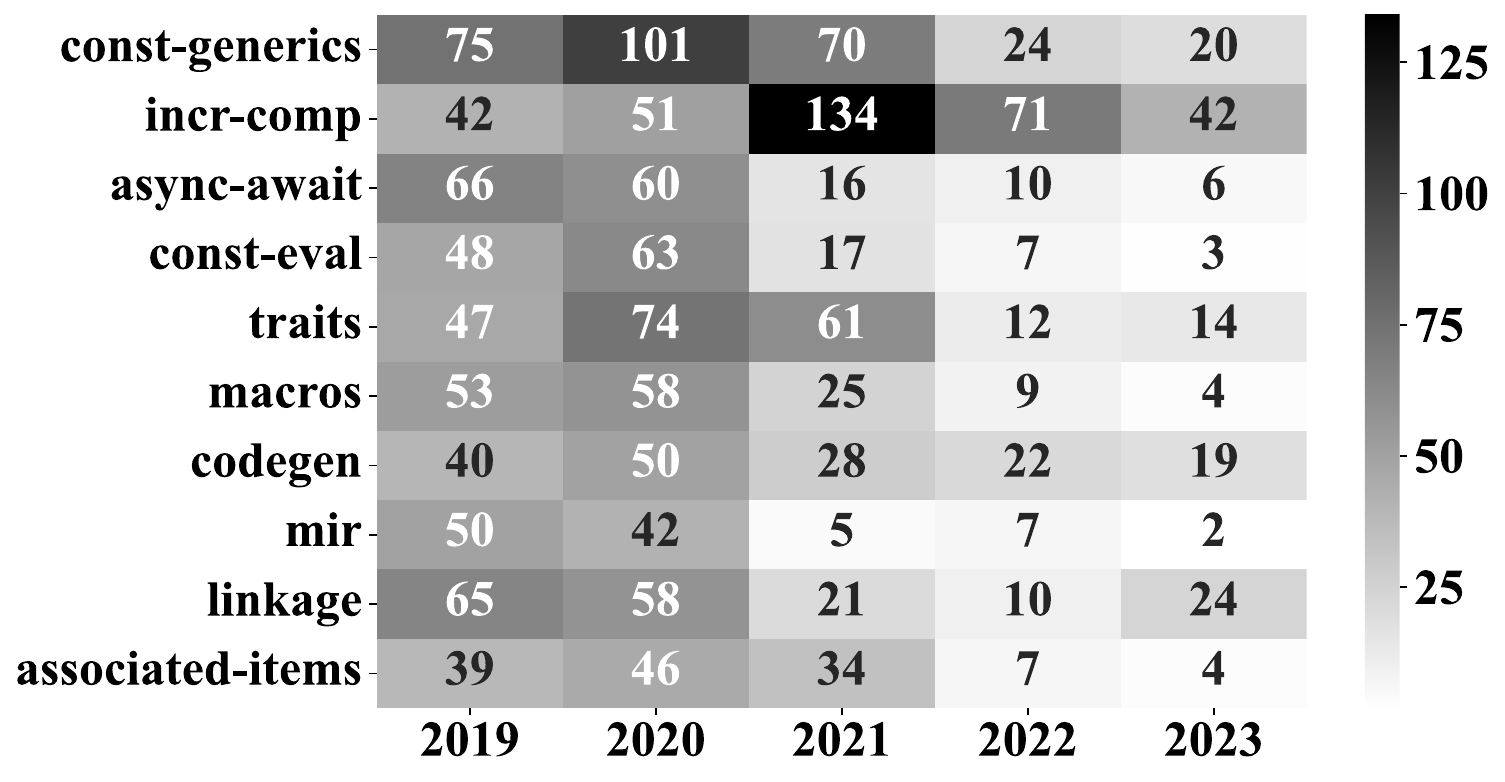}
    \caption{Evolution of Bug Reports in \textit{rustc} Components from January 2019 to August 2023: A Heatmap Analysis}
    \label{fig:heatmap}
\vspace{-0.3cm}
\end{figure}
A web crawler was developed to collect code snippets and label information from bug issues in \textit{rustc}. As we focus only on compiler-related bugs, during the collection phase, we specifically targeted issues with the ``\texttt{C-bug}" and ``\texttt{T-compiler}" labels, resulting in a total of 7,474 issues (of which 5,242 are closed). We utilized BeautifulSoup~\cite{butterfly}, which is a Python library for parsing HTML and XML documents, to extract label information for each bug issue. Through regular expression matching, we extracted a total of 6,088 code snippets (including those mentioned in comments) from 3,819 resolved bug issues.

Our analysis shows it takes about 209 days on average to resolve a Rust compiler bug, with some taking over a decade (e.g., \href{https://github.com/rust-lang/rust/issues/10186}{\textcolor{mypurple}{\textbf{\#Issue10186}}}). This highlights the significant delay and complexity in fixing Rust compiler issues, which can persist across versions.

\subsubsection{Findings}
\label{subsec:findings}
By statistically analyzing the code snippets and labels extracted from bug issues that historically triggered compiler errors, we have obtained two crucial observations.

\ding{192} \textbf{The code that historically triggered compiler bugs contains a rich set of features, involving various components of previously released compilers.} Based on our analysis of code extracted from issues, we found that 31\% (2016 out of 6088) of the code contains the declaration ``\greenbf{\#![feature(...)]}". 
Of the code snippets containing feature declarations, each code segment contains 1.15 feature declarations on average, with the maximum combination containing 8 features.
 These explicit  feature declarations in the code can enable or disable compiler pipelines, which is a significant factor leading to compiler bugs. 
By examining the labels provided by developers for bug issues, we further analyzed the compiler components involved in these code samples. Figure \ref{fig:heatmap} presents compiler components' top 10 ranked label information associated with confirmed \textit{rustc} issue reports from January 2019 to August 2023. We discovered that the compiler code triggering historical bugs involves various components, such as const generics, incremental compilation, async/await, const evaluation, traits system, macros, code generation, intermediate representation, linkage, and associated item handling. Notably, const generics and incremental compilation, two components of \textit{rustc}, have consistently shown a higher number of reported issues, indicating that they are components that are more prone to trigger bugs. 
 From Figure \ref{fig:heatmap}, it can be observed that as \textit{rustc} continues to iterate and developers improve the functionality of its components, there is a relative decrease in the number of newly added bug issues for the internal components of the compiler.
However, there still remains room for testing and exploration. According to the observations by Zhong~\cite{HistIssue} and Deng~\cite{2023arXiv230402014D}, code that historically triggers compiler defects can explore edge cases in the compiler.

$\bigstar$ \textit{\textbf{Conclusion}}: Therefore, due to the ongoing development of certain features declared in ``\greenbf{\#![feature(...)]}", they are not sufficiently stable and may contain potential bugs. When constructing test cases for the Rust compiler, special consideration should be given to these features.
\begin{figure}
    \centering
    \includegraphics[width=0.5\textwidth]{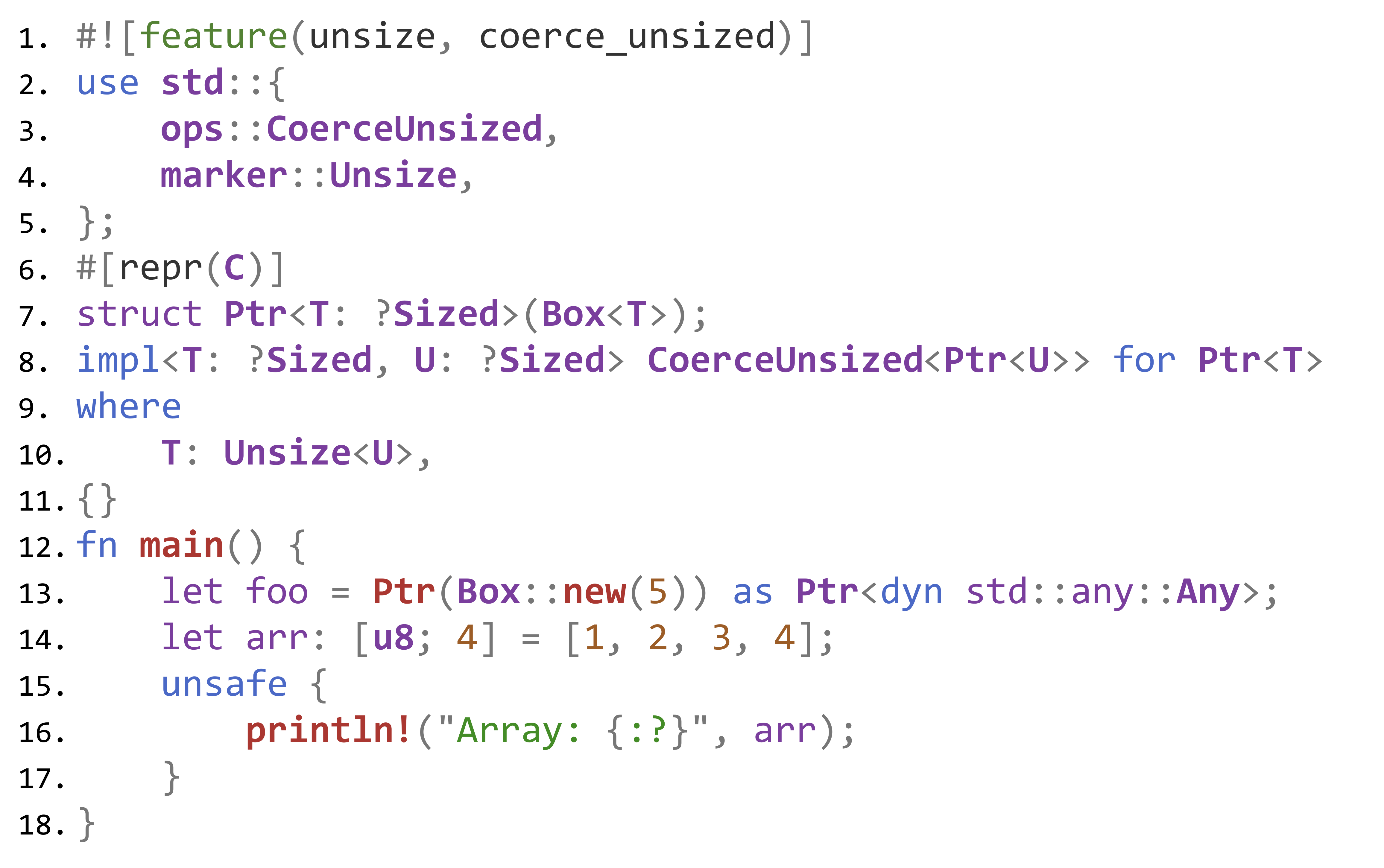}
    \caption{An Example: Test Code from \textit{rustc} ~\href{https://github.com/rust-lang/rust/issues/98322}{\textbf{\textcolor{mypurple}{Issue\#98322}}}}
    \label{fig:exampleCode}
\vspace{-0.3cm}
\end{figure}

\ding{193} \textbf{The majority of test code that triggers the Rust compiler exhibits abundant usage of bracket structures.} These bracket structures, including ``\texttt{()}", ``\texttt{\{\}}", ``\texttt{$[]$}", and ``\texttt{<>}", play a crucial role in defining the syntax and semantics of Rust code. They are used for various purposes, including function and method invocations, control flow statements, data structure definitions, and pattern matching, among others. The widespread use of bracket structures indicates that Rust's syntax heavily relies on these constructs to provide expressive and flexible language features. Taking Figure~\ref{fig:exampleCode} as an example, \texttt{\#![feature(unsize,coerce\_unsized)]} and \texttt{\#$[$repr(C)$]$} are both attribute declarations, where the square brackets $[]$ denote attributes. \texttt{\#![feature(unsize,coerce\_unsized)]} enables two experimental features, namely \texttt{unsize} and \texttt{coerce\_unsized}. \texttt{\#$[$repr(C)$]$} is applied to the \texttt{Ptr} structure to ensure its memory layout is compatible with C language structs.
The syntax \texttt{<T: ?Sized>} represents a generic constraint, known as a generic bound, which limits the behavior of the type parameter T. In this context, it restricts T to be a type that may or may not have a known size.
The expression \texttt{$[$u8; 4$]$} initializes an array of size 4 with elements of type u8.
The \texttt{unsafe \{\}} block is used to indicate that the enclosed code contains operations that bypass Rust's memory and safety checks. It allows certain operations that are not considered safe by the Rust compiler. The purpose of using \texttt{unsafe \{\}} is to circumvent some of Rust's safety guarantees and gain additional flexibility or performance, but it requires the programmer to ensure correctness and safety manually.

$\bigstar$ \textit{\textbf{Conclusion}}: The presence of diverse bracket structures in code highlights compilers' need to handle various grammar cases accurately. When performing code generation tasks for Rust test code, it can be beneficial to incorporate code snippets featuring a range of bracket structures as contextual input to guide the LLMs. This approach ensures coverage of different syntactic combinations and boundary scenarios, thereby facilitating the generation of comprehensive test code that maximizes coverage across various compiler components.

%% file: content/approach.tex
\begin{figure*}
    \centering
    \includegraphics[width=0.78\textwidth]{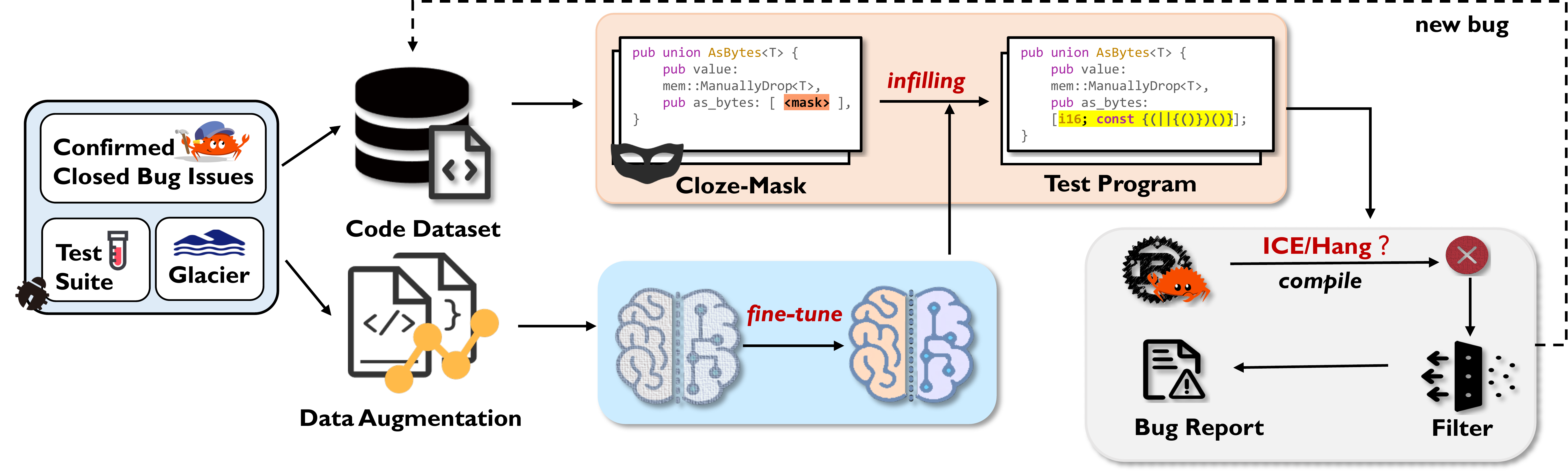}
    \caption{The Overview of \tool~}
    \label{fig:overview}
\vspace{-0.3cm}
\end{figure*}

In this section, we present our method, \tool, designed to discover new bugs in the Rust compiler. Our approach harnesses the power of LLM and a curated collection of code snippets that have historically exposed bugs in the Rust compiler.

We employ a \texttt{clozeMask} strategy (Section \ref{sec:step3}) to synthesize new test cases. 
Our empirical analysis (Section\ref{subsec:findings}) shows the test code historically triggering Rust compiler bugs exhibits rich language features. Therefore, the core assumption of \tool~is code snippets triggering historical bugs can explore vulnerable components and logic branches of the compiler. Leveraging these diverse code snippets, we can generate novel test cases to effectively fuzz the Rust compiler. Previous work in this direction required significant manual effort, where experienced developers had to devise a set of tailored rules and constraints specific to the compiler under test. Moreover, directly generalizing these approaches to testing the Rust compiler was challenging. In contrast, recent Large Language Model (LLM) advancements provide a natural, generalizable, and fully automated solution. Modern LLMs easily ingest historical programs via prompting or fine-tuning, generating programs similar to historical ones, effectively utilizing their code components.
Figure \ref{fig:overview} illustrates an overview of \tool. We systematically mine bug reports from the \textit{rustc} code repository to collect code snippets historically triggering bugs in the Rust compiler. Our original code corpus includes the official Rust test suite~\cite{testsuite}. Furthermore, the Rust-lang team provides the Glacier~\cite{glacier}, which contains Rust code snippets and shell execution scripts that historically triggered internal compiler errors (ICEs). We also include these code snippets in our dataset. Leveraging these code snippets that trigger historical compiler bugs, \tool~ can generate new rust programs to effectively fuzz the Rust compiler, aiming to discover new bugs. 

In our research, the \tool~follows a four-step workflow, which can be outlined as follows:
\begin{itemize}[leftmargin=*]
    \item \textbf{Step-1. }\textit{\textbf{Dataset construction}}. This step involves web scraping to collect data for building a code dataset for Step-3 and a dataset with augmented data for Step-2. (Section \ref{sec:step1})
    \item \textbf{Step-2. }\textit{\textbf{Fine-tuning}}. The LLM is fine-tuned using historical test code that triggered bugs. This process enables the LLM to learn the characteristics of test code that historically triggered bugs. (Section \ref{sec:step2})
    \item \textbf{Step-3. }\textit{\textbf{ClozeMask}}. 
    A code snippet is randomly selected from the code dataset, and the positions of bracket structures within the snippet are identified using a depth-first search (DFS) strategy. The content enclosed within the brackets is substituted with a ``\mybox[mymask]{mask}'' label, and an LLM is used to populate the masked portion, thereby generating new test cases. (Section \ref{sec:step3})
    \item \textbf{Step-4. }\textit{\textbf{Oracle and Duplicated Bug Filtering}}. The test oracle is used to determine whether a bug is present. If a bug is identified, a bug deduplication process is performed. For new bugs, the test case is added to the code dataset for dataset updates. (Section \ref{sec:step4})
\end{itemize}

In the following sections, we will provide a more comprehensive explanation of each step in \tool.

\subsection {Dataset Construction}
\label{sec:step1}
\subsubsection{Building the Code dataset}
Firstly, we implemented an HTML crawler to collect all bug issues related to \textit{rustc} and extracted Rust code snippets for inclusion in our dataset. Specifically, we selected closed issues tagged with ``\texttt{T-compiler}" and ``\texttt{C-bug}" up until August 13, 2023. We used regular expressions to extract Rust code segments from these issues. Additionally, to supplement our dataset, we extracted suitable \textit{.rs} files from the official Rust test suite and the Glacier repository, as well as Rust code from \textit{.sh} files. These additions were incorporated into our code dataset. Importantly, to avoid interfering with experimental results,  we removed code snippets from Glacier and the test suite that could trigger compiler bugs originally.

\subsubsection{Data Augmentation}
To fine-tune the LLM, it is necessary to augment the Rust code corpus in our raw code dataset. According to Dong et al.'s empirical research~\cite{code.data.aug}, data augmentation techniques that slightly disrupt source code syntax, such as random deletion (RD) and random swapping (RS), have proven effective in enhancing PL models pre-training. Therefore, we employ a random approach with a probability of $p$ to delete code tokens at the code segment level in the original dataset. We manually set the value of $p$ to 0.2, as this allows us to retain the original semantic meaning of the code as much as possible when implementing the RD strategy. Additionally, we apply RS as a data augmentation technique for code segments in the original dataset. Specifically, for a given test program, we randomly select two statements and interchange their positions, repeating $n$ times. The value of $n$ is determined based on the number of tokens in the code. The original dataset contains 25248 code snippets. Through data augmentation, we expand the dataset to 100000 code snippets.
\normalem
\begin{algorithm}[htp]
     \SetKwFunction{FnclozeMask}{cloze}
     \SetKwFunction{FngetMaskedPositions}{getMaskedPositions}
     \SetKwFunction{FnInfilling}{Infilling}
	
	\SetKwProg{Fn}{Function}{}{end}
	\SetKwProg{For}{for}{}{end}
	\SetKwProg{Fp}{Procedure}{}{end}
     \SetKwProg{Else}{else}{}{end}
	\caption{\small Main procedure of \texttt{clozeMask}}
	\label{ago:cloze}
	\DontPrintSemicolon
    \small

	\Fn{\FnclozeMask{$Code_{raw}$}}{  
    $Pos_{mk}$ $\leftarrow$ \FngetMaskedPositions{$Code_{raw}$} \label{algo:getmaskedpositions}  \;
    \Comment{\textcolor{mycommentcolor}{DFS is employed to extract the positions of}}
    \Comment{\textcolor{mycommentcolor}{bracket structures}}
    $Code_{mk} \gets \texttt{[ ]}$\;
    \BlankLine
    \For{$\text{pos}$ \textbf{in} $Pos_{mk}$}{
    $Code_{mk} \gets Code_{raw}.\text{replace}(\text{pos}, \texttt{"\mybox[mymask]{mask}"})$ \;
    $Codes_{mk}.\text{append}(Code_{mk})$
    }
    \Return{$Codes_{mk}$}
    }

  \Fn{\FnInfilling($Code_{raw}$,$time_{max}$,$t$)}{
    $Codes_{new} \gets \texttt{[ ]}$\;
    $Codes_{mk} \gets \FnclozeMask(Code_{raw})$\;
    \BlankLine
    \For{$Code_{mk}$ \textbf{in} $Codes_{mk}$}{
        \eIf{\text{isSpecialMasked}($Code_{mk}$)}{
            \Comment{\textcolor{mycommentcolor}{bracket pairs in "feature" block}}
            \For{$i \gets 0$ \textbf{to} $time_{max}$}{
                        $t_r \gets \text{randomFloat}(0, 1)$\;
                        $Code_{new} \gets \text{codeInfilling}(Code_{mk}, t_r)$\;
                        \If{$Code_{new} \neq Code_{raw}$}{
                        $Codes_{new}.\text{append}(Code_{new})$\;
                        }
            }
        }{
            $Code_{new}  \gets \text{codeInfilling}(Code_{mk},t$)\;
                \If{$Code_{new} \neq Code_{raw}$}{
                $Codes_{new}.\text{append}(Code_{new})$\;
                }
        }
    }
  \Return{$Codes_{new}$}
  }

\end{algorithm}
\normalem
  \vspace{-0.5cm}
\subsection{Fine-tuning}
\label{sec:step2}
We used fine-tuning to improve the code generation task for Rust compiler bugs. Fine-tuning is a transfer learning technique where a pre-trained model is further trained for specific task requirements. We used the Incoder model~\cite{incoder}, an open-source LLM supporting Rust code completion with fewer parameters. We performed fine-tuning by treating code snippets as training samples, encoded using Byte Pair Encoding (BPE)~\cite{BPE1,GPT2}. The last few layers were trained to maximize the probability of predicting the next token. The Adam optimizer~\cite{Adam} expedited convergence. Retraining the Incoder model took over 48 hours, resulting in a language model capable of generating test programs from seed code snippets.
}
\subsection{ClozeMask}
\label{sec:step3}
To generate Rust test programs, we employed the \texttt{clozeMask} method to transform test code from an existing code dataset. The \texttt{clozeMask} algorithm consists of two processes: \texttt{cloze} and \texttt{infilling}, illustrated in Algorithm \ref{ago:cloze}.

\subsubsection{Cloze}

The provided Rust code utilizes a stack and recursion to preserve the occurrences of bracket pairs (including ``\texttt{()}", ``\texttt{\{\}}", ``\texttt{$[]$}", and ``\texttt{<>}") within the given code text. By traversing each bracket pair's position in the stack, the original code snippets are replaced with the ``\text{\mybox[mymask]{mask}}", resulting in various code cloze frameworks for completion. This approach enables the creation of different code fill-in-the-blank templates by systematically identifying and masking the bracket pairs within the code text.

\subsubsection{Infilling}
Masked codes generated from the traversal of the original code are iteratively populated. In cases where the insertion point of the mask in a masked code is considered special, additional fillings are performed with varying frequencies and temperature parameters. This approach aims to obtain test codes that are more likely to trigger compiler bugs. If the populated code remains identical to the original code after filling, the generated result is discarded to ensure high-quality of test cases.

\subsection{Oracle and Duplicated Bug Filtering}
\label{sec:step4}
We assess the Rust compiler's performance using generated outputs with general testing oracles. Additionally, we develop a methodology for identifying repetitive bugs unique to the Rust compiler. 
\subsubsection{Oracle}
To evaluate the generated outputs, we use two testing oracles: ICE and Hang, specifically designed for testing the Rust compiler. The ICE issue in the Rust compiler refers to ``Internal Compiler Error". When the compiler encounters an unhandled or unexpected situation, it crashes and displays an ICE error message. The Hang issue in the Rust compiler refers to situations where the compilation process becomes stuck or unresponsive. When the compiler gets stuck in an infinite loop, consumes resources for an extended period, or cannot proceed with execution, a Hang issue occurs. Both types of compiler bugs can be detected during compilation by examining stack information and observing compilation duration, without requiring a comparative compiler. 

\subsubsection{Duplicated Bug Filtering}
We collect stack trace information for test cases triggering ICE errors into a bug dataset and gather \textit{time-passes} information for test cases triggering Hang issues into the bug dataset as well. For test cases triggering bugs, we search the bug dataset. If the stack trace information or \textit{time-passes} information matches an entry in the bug dataset, we discard the test case. If there is no match, we add the corresponding stack trace information or \textit{time-passes} information to the bug dataset, update it, and mark the test case as ``interesting" to generate a bug report. We store the test case in the code dataset to enrich it for subsequent iterations. This process eliminates duplicate bugs' impact on the method and ensures new bug detection.

%% file: content/evaluation.tex
\newcommand{\cis}{C}
\newcommand{\nis}{\overline{C}}

In this section, we conduct extensive evaluations to investigate the effectiveness of our method. 
Specifically, the experiments aim to answer the following research questions:

\begin{itemize}[leftmargin=*]
    \item \textbf{RQ1:} What is \tool's capability in bug hunting? (Section~\ref{sec:rq1})
    \item \textbf{RQ2:} How does \tool~compare against existing fuzzers in terms of effectiveness? (Section~\ref{sec:rq2}) 
    \item \textbf{RQ3:} How do the key components of \tool~ contribute to its effectiveness? (Section~\ref{sec:rq3})
\end{itemize}

\subsection{Evaluation Setup}
\label{subsec:EvaluationSetup}

\mypara{Target Compilers.}
We applied the \tool~to two compilers, \textit{rustc} and \textit{mrustc}, as indicated in Table \ref{tbl:Compilers}. Of these two compilers, \textit{rustc} represents the official Rust language compiler, while \textit{mrustc} is an individually developed Rust compiler that currently supports a limited subset of Rust syntax. 
Hence, our experimental investigations primarily center around \textit{rustc}, which holds dual importance as the sole compiler officially recognized by the Rust language development team and for its extensive support of the latest and most comprehensive Rust language syntax.
\begin{table}[h]
\setlength{\abovecaptionskip}{0cm} 
\caption{Target compilers we have tested.}
\label{tbl:Compilers}
\setlength{\tabcolsep}{6mm}{
\begin{center}
\begin{tabular}{lll}
\hline
\rowcolor[HTML]{FFFFFF} 
\textbf{Compilers}                                                            & \textbf{Versions} & \textbf{Build No.} \\ \hline
\rowcolor[HTML]{FFFFFF} 
\multicolumn{1}{l|}{\cellcolor[HTML]{FFFFFF}}                                 & v1.74-stable      & 79e9716c9          \\
\rowcolor[HTML]{FFFFFF} 
\multicolumn{1}{l|}{\cellcolor[HTML]{FFFFFF}}                                 & v1.73-stable      & cc66ad468          \\
\rowcolor[HTML]{FFFFFF} 
\multicolumn{1}{l|}{\multirow{-3}{*}{\cellcolor[HTML]{FFFFFF}\textit{rustc}}} & v1.72-stable      & d5c2e9c34          \\ \hline
\rowcolor[HTML]{EFEFEF} 
\multicolumn{1}{l|}{\cellcolor[HTML]{EFEFEF}\textit{mrustc}}                  & v1.29.100         & 4c7e8171           \\ \hline
\end{tabular}
\end{center}
}
\vspace{-0.3cm}
\end{table}
\begin{itemize}[leftmargin=*]
    \item \textbf{\textit{rustc},} the official compiler of the Rust programming language, is  developed in Rust itself. \textit{rustc} follows a ``self-compilation" methodology and employs the LLVM framework as its backend architecture.
    \item \textbf{\textit{mrustc},}  a Rust compiler variant developed by an individual programmer using C++, represents a ``simplified" version of the compiler that supports a subset of Rust syntax.
\end{itemize}

\mypara{Bug Type.}
To ensure the reliability of the Rust compiler, our evaluation primarily focuses on the bugs mentioned in Section \ref{sec:step4}, namely ICE and Hang. 
\begin{itemize}[leftmargin=*]
    \item \textbf{ICE:} When the \textit{rustc} compiler crashes during compilation due to an unexpected trigger, resulting in the output ``\texttt{internal compiler error}" or ``\texttt{compiler unexpectedly panicked}", we consider it an ICE issue. Similarly, if the \textit{mrustc} compiler interrupts compilation and outputs ``\texttt{BUG}" with ``\texttt{core dump}", it is also a potential ICE problem.
    \item \textbf{Hang:}  The ``Hang" issue in compilers refers to a situation where the compiler becomes unresponsive or stuck during the compilation process, causing it to stop making progress or producing any output. 
    According to industry conventions, a compiler is considered to have a hang issue if its compilation time exceeds 60s. Taking into account that the Rust compiler tends to have slower compilation speeds compared to traditional compilers~\cite{slowRust}, we have relaxed this timeout threshold $T$ and set it to 180s.
    The 100\% confirmation rate of bug issues related to Hang, as reported by us, validates the reasonableness of setting threshold $T$.
\end{itemize}

\mypara{Competitive Baselines.} 
We selected RustSmith~\cite{Rustsmith}, Rustlantis~\cite{rustlantis.pro}, and SPE~\cite{spe} as our baseline methods. RustSmith and Rustlantis are open-source generative fuzzers, and we utilized their latest versions, RustSmith-1.30.0 and Rustlantis-0.1.0 in our experiments. SPE is a well-known compiler testing case generation method based on Skeletal Programs Enumeration. It was originally designed as a mutation-based fuzzer for C compilers. We have re-implemented SPE to meet the testing requirements of the Rust compiler. For a fair comparison, the seed programs used by SPE are identical to those used by \tool. Specifically, we extract the variables from the seed programs to obtain a variable set $V$, and the program skeleton $P$ after the variable extraction. We then rearrange the variables and fill them back into $P$ to generate new test programs. Since the rearrangement of the variable set $V$ is a full permutation problem, we simplify the process by filtering the full permutations for each skeleton $P$ (if the full permutation result exceeds a threshold $T_{\theta}$=64, we randomly select 32 from the full permutation result) before refilling the skeleton. The proportion of cases where the full permutation results of the variable set $V$ exceed the threshold $T_{\theta}$ is 4.84\%.

\mypara{Implementation and Evaluation Platforms.}
Our program generator is developed using PyTorch v2.1.0 and CUDA version 12.1. The implementation combines Rust and Python languages, with the core code comprising approximately 500 lines.
The evaluation platform we employed is a multi-core server equipped with a 20-core CPU, 4 Tesla V100-SXM2-32GB GPUs, and 216GB of RAM. The server runs on the Ubuntu 18.04 operating system with Linux kernel version 4.15. All LLMs are executed on this server using GPUs. 

\subsection{RQ1: Effectiveness}
\label{sec:rq1}
From August 2023 to November 2023, we extensively tested \textit{rustc} and \textit{mrustc} using \tool. To ensure that all detected bugs were new, we always ran \tool~on the latest trunk of the compilers for testing purposes.

\begin{table}
    \caption{Bugs found by \tool{}~ in \textit{rustc} and \textit{mrustc}.}
    \label{tbl:BugReport}
    \centering
    \renewcommand{\arraystretch}{1.25}
    \setlength{\tabcolsep}{6pt}
    \small
    \begin{tabular}{lrr|rr|c}
        \toprule
        \multirow{2}{*}{\textbf{Status}} & \multicolumn{2}{c|}{\textbf{rustc}} & \multicolumn{2}{c|}{\textbf{mrustc}} & \multirow{2}{*}{\textbf{Total}} \\
        & \textit{ICE} & \textit{Hang} & \textit{ICE} & \textit{Hang} & \\
        \midrule
        Reported & \rireported & \rhreported & \mireported & \mhreported & \reported \\
        Confirmed & \riconfirmed & \rhconfirmed & \miconfirmed & \mhconfirmed & \confirmed \\
        Fixed & \rifixed & \rhfixed & \mifixed & \mhfixed & \fixed \\
        Duplicate & \riduplicate & \rhduplicate & \miduplicate & \mhduplicate & \duplicate \\
        Won't fix & \riwontfix & \rhwontfix & \miwontfix & \mhwontfix & \wontfix \\
        \bottomrule
    \end{tabular}
    \vspace{-0.3cm}
\end{table}

\mypara{Bug Count.}
In general,~\tool~generated over 1000 potential bug-triggering instances, and all confirmed bugs were triggered by the Rust compiler's basic optimization level, \texttt{-O0}. Following the bug deduplication principle mentioned in Section~\ref{sec:step4}, we reported a total of \reported~bugs involving \textit{rustc} and \textit{mrustc}, out of which \confirmed~bugs have been confirmed or fixed by the developers, as detailed in Table \ref{tbl:BugReport}. It is worth noting that all the confirmed bugs were previously unknown and could not be triggered by any historically known bug-triggering code. In total, these confirmed bugs resulted in 12 ICEs. Additionally, \tool~revealed 15 confirmed Hang issues in \textit{rustc}, which are significant vulnerabilities in the Rust compiler. Some of the vulnerabilities that developers deemed ``\textit{won't fixed}" were primarily due to the test cases utilizing internal features of the compiler that are not publicly accessible (\href{https://github.com/rust-lang/rust/issues/116783}{\textcolor{mypurple}{\textbf{\#116783}}}, \href{https://github.com/rust-lang/rust/issues/116784}{\textcolor{mypurple}{\textbf{\#116784}}}). This is mainly because we used certain code snippets from the testsuite that are not exposed to external developers.

\mypara{Affected compiler components.}
To assess the performance of \tool~in covering different components of the compiler, we classified the bugs that have been confirmed with their root causes based on developers' feedback and labeled information. Table \ref{tbl:components} presents the categorization of these bugs. Taking \textit{rustc} as an example, we found that the bugs identified by \tool~span across multiple components of the compiler, including \textit{parser}, \textit{macros}, \textit{associated-items}, \textit{traits}, and \textit{lifetimes}. This observation indicates that \tool~has the potential to thoroughly explore different components and unveil compiler bugs. Additionally, we discovered that out of all the confirmed bugs, 8 were attributed to features, with seven being ICE issues and one being a Hang issue. Notably, the ``\greenbf{generic\_const\_exprs}" feature was responsible for the majority of these errors, occurring in 8 out of 5 instances. 
This finding suggests that increasing the number of mutations involving ``\greenbf{\#feature[…]}" during the \texttt{clozeMask} stage is more likely to generate test cases that expose compiler errors, corroborating our pre-study in Section~\ref{subsec:preStudy}.

\begin{table*}[]
\setlength{\abovecaptionskip}{0cm} 
\caption{Compiler Components Implicated in Bugs Confirmed with Root Causes by Developers.}
\label{tbl:components}
\setlength{\tabcolsep}{6mm}{
\begin{center}
\begin{tabular}{
>{\columncolor[HTML]{FFFFFF}}c 
>{\columncolor[HTML]{FFFFFF}}l 
>{\columncolor[HTML]{FFFFFF}}l 
>{\columncolor[HTML]{FFFFFF}}c }
\hline
\textbf{Compiler}                                        & \textbf{Component}            & \textbf{Brief Description}                                                                  & \textbf{Bugs}            \\ \hline
\cellcolor[HTML]{FFFFFF}                                 & parser                        & The parsing of Rust source code to an AST.                                                  & 3                        \\
\cellcolor[HTML]{FFFFFF}                                 & macros                        & All kinds of macros (\textit{custom derive, macro\_rules!, proc macros, ..})                         & 1                        \\
\cellcolor[HTML]{FFFFFF}                                 & associated items              & Associated items such as associated types and consts.                                       & 2                        \\
\cellcolor[HTML]{FFFFFF}                                 & {\color[HTML]{333333} traits} & Trait system                                                                                & {\color[HTML]{333333} 2} \\
\cellcolor[HTML]{FFFFFF}                                 & lifetime                      & lifetime related                                                                            & 1                        \\
\cellcolor[HTML]{FFFFFF}                                 & const-genetics                & const generics (parameters and arguments)                                                   & 1                        \\
\multirow{-7}{*}{\cellcolor[HTML]{FFFFFF}\textit{rustc}} & feature                       & when including `\greenbf{\#feature{[}...{]}}' in source code                                          & 8                        \\ \hline
\textit{mrustc}                                                   & genetic-type-check            & generic type was copied to a scope with no generics                                         & 1                        \\
                                                         & macros                        & \cellcolor[HTML]{FFFFFF}All kinds of macros (\textit{custom derive, macro\_rules!, proc macros, ..}) & 1                        \\ \hline
\end{tabular}
\end{center}
}
\vspace{-0.4cm}
\end{table*}

\begin{figure}
    \centering
    \includegraphics[width=0.40\textwidth]{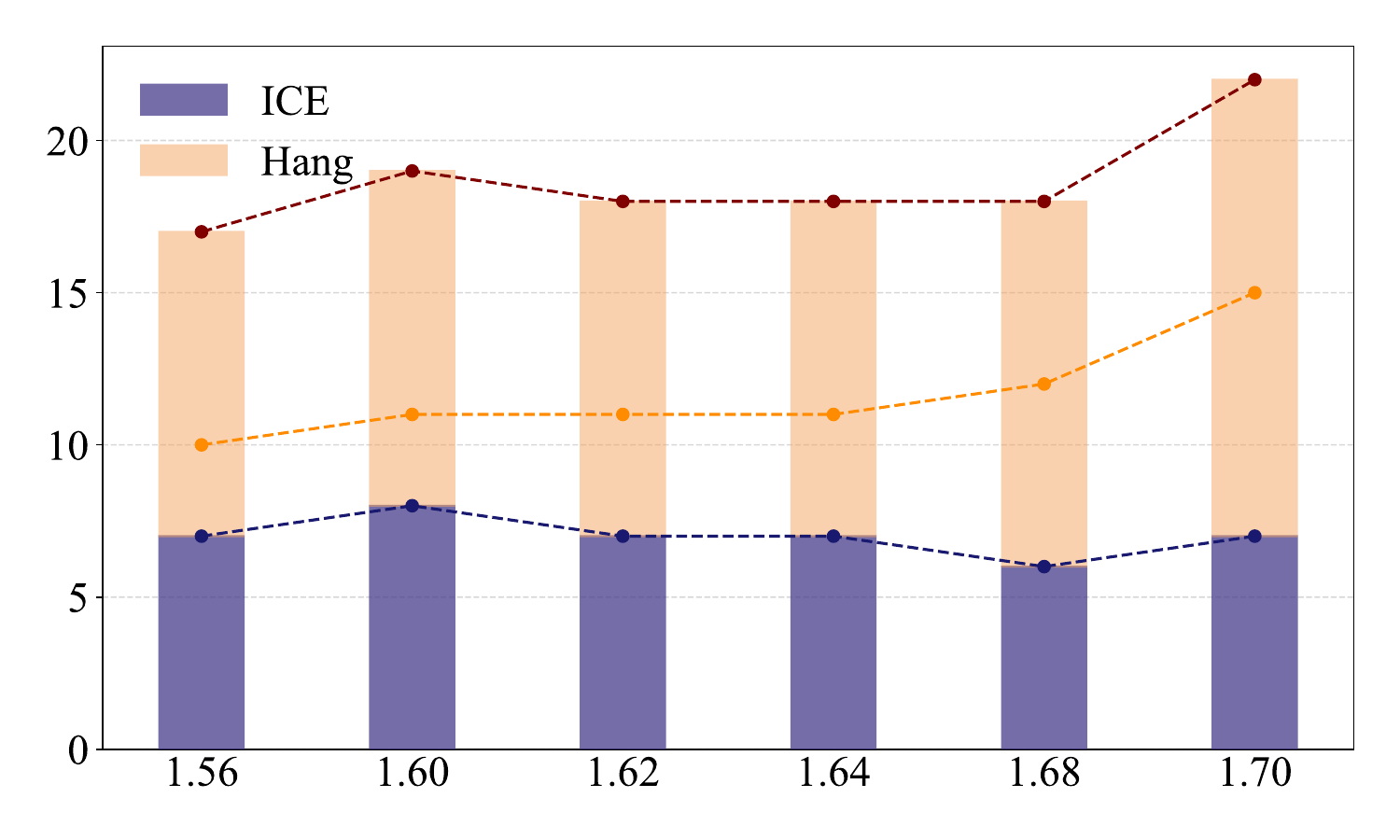}
    \caption{  Confirmed bugs that affect the corresponding release versions of \textit{rustc}}
    \label{fig:regression}
\end{figure}

\mypara{Significance.} Many bugs discovered by our tool have received prompt feedback and active discussions from the developers. For example, in the case of \href{https://github.com/rust-lang/rust/issues/116647}{\textcolor{mypurple}{\textbf{Issue\#116647}}} reported on \textit{rustc}, the developers acknowledged the issue and provided a comment within one day, stating, ``\textit{Interestingly, the time gets significantly worse if the trailing comma for the last line is removed.}"
Furthermore, to gain a better understanding of the significance of identified bugs, we conducted a study on how the confirmed bugs discovered by \tool~impact the historical versions of \textit{rustc}. Specifically, we input the submitted triggering test code, which causes new bugs, into a series of official release versions of \textit{rustc} and observed whether any abnormal results occurred. The formal stable versions we selected for this study were 1.56 (released on November 1, 2021),1.60, 1.62, 1.64, 1.68, and 1.70 (released on May 31, 2023), which were the versions available before our experiment. 
These versions have undergone extensive testing with various tools, such as RustSmith and Rustlantis. As shown in Figure \ref{fig:regression}, we identified 17 bugs that initially appeared in \textit{rustc} stable version 1.56. In other words, they evaded detection by other fuzzers and remained latent in \textit{rustc} for over two years. Additionally, we discovered that some test cases triggered ICE bugs in certain versions but manifested as Hang bugs in other versions. This observation signifies that the test cases produced by our tool effectively investigate various edge cases of the Rust compiler. Furthermore, as subsequent versions of the compiler are developed, developers may encounter difficulties in effectively addressing these intricate edge cases.
\begin{figure}[t]
    \centering
    \includegraphics[width=0.4\textwidth]{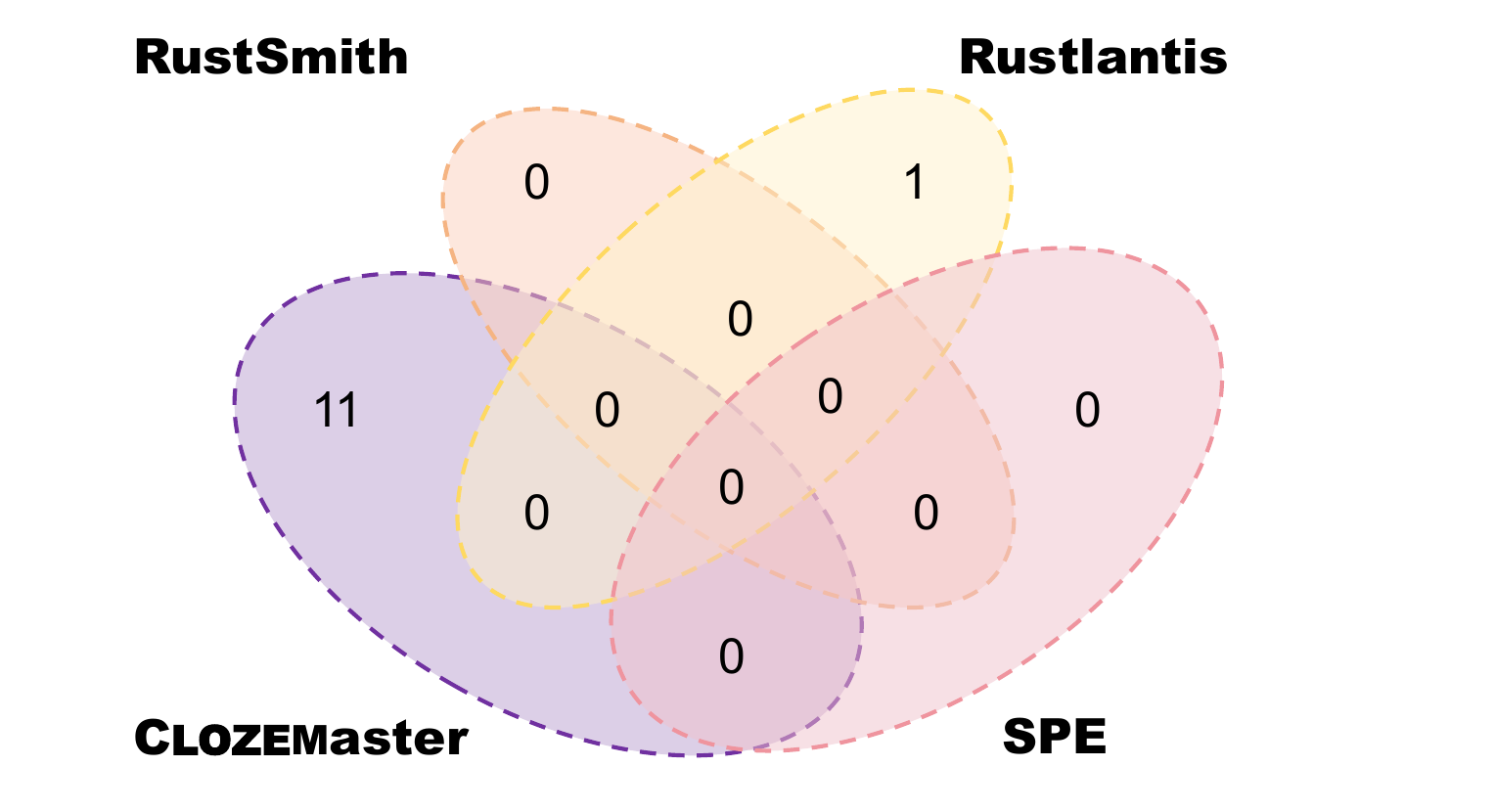}
    \caption{The distribution of unique bugs detected by \tool~ and the baselines on the stable version 1.73 of \textit{rustc}.}
    \label{fig:vnn}
\end{figure}
\begin{tcolorbox}[notitle,boxrule=0pt,colback=yellow!10,colframe=blue!20] 
\textbf{Summary on Effectiveness}: \tool~ possesses the capability to synthesize effective test cases that reveal novel bugs within the Rust compiler. These test cases expose bugs across various components of the Rust compiler, including those that have remained undiscovered in previous versions by developers.
\end{tcolorbox}

\subsection{RQ2: Comparison with Existing Techniques}
\label{sec:rq2}
In this research question (RQ2), we investigate whether our approach improves upon existing techniques, including Rustsmith, Rustlantis, and SPE. 
In the experiment, RustSmith and Rustlantis utilized randomly generated integer values as random seeds to guide the generation of their test cases. The SPE method, however, requires existing test programs as input. To ensure the fairness of the experiment, the original seed program dataset from \tool~was used as the input for SPE.
Compared to the conventional generation strategies employed by other fuzzers, the results demonstrate the effectiveness of our proposed framework, which leverages LLM as a cloze master to generate high-quality Rust test cases.
Here, we primarily focus on comparing their bug-finding capabilities and the achieved code coverage across different testing tools.

\begin{table}[]
\setlength{\abovecaptionskip}{0cm} 
\caption{Code coverage achieved by different fuzzers. The
highest coverage is highlighted. The baseline fuzzers are bolded.}
\label{tbl:coverage}
\renewcommand{\arraystretch}{1.25}
\setlength{\tabcolsep}{4mm}{
\begin{center}
\begin{tabular}{l|l}
\hline
Techniques                    & \textit{Code Coverage} \\ \hline
\textbf{RustSmith}  &  32.84\%         \\
\textbf{Rustlantis} &  29.55\%      \\
\textbf{SPE}        & 62.02\%      \\
\rowcolor[HTML]{FFCCC9} 
\tool          & \textbf{64.34}\%            \\ \hline
\end{tabular}
\end{center}
}
\vspace{-0.4cm}
\end{table}

\subsubsection{\textbf{Bug-finding Capability}}
To evaluate the bug-finding capability of \tool, we re-ran \tool~and the baseline tools for 24 hours. 
Besides, to mitigate the potential influence of data leakage, we selected the stable version 1.73 of \textit{rustc} (released on October 3, 2023) as the test subject in our experiment. The bug-triggering inputs used for training are primarily reported and resolved prior to the release of this version, thereby avoiding the data leakage problem. 
The results are presented in Figure \ref{fig:vnn}. 
We observed that \tool~identified the highest number of bugs, totaling 11, compared to the other tools. Furthermore, we noted that there was minimal overlap between the bugs detected by \tool~and the other tools. 
Furthermore, we found that template-based test program generators such as RustSmith and Rustlantis can limit the diversity of the test programs they produce. For instance, the test programs generated by RustSmith often exhibit similar macro declarations and expression structures, while Rustlantis utilizes a singular feature declaration approach along with pattern matching.
This is why they were able to find a relatively large number of bugs in the early versions of \textit{rustc}, but as \textit{rustc} has been continuously improved, these two methods have difficulty detecting new bugs in the newer versions of \textit{rustc}. 
In contrast, SPE exhibits a strong dependence on the positioning of variables within the seed programs, and its mutation process is restricted to variable replacement alone. Conversely, the \tool~supports more flexible mutations enabled by LLM that can better comprehend the contextual semantics of the code.
This indicates that \tool~exhibits strong complementarity with existing fuzzers and demonstrates a higher degree of bug-finding capability.

\subsubsection{\textbf{Code Coverage}}
We conducted a comparative analysis of code coverage on the Rust compiler between baseline methods and \tool, generating an equal number of 10,000 test cases for evaluation to ensure a fair comparison. These test cases were also executed using the stable version 1.73 of \textit{rustc}. 
We employ an instrumentation-based approach within the Rust compiler, \textit{rustc}, to collect code coverage metrics for test cases generated by different methods.
Individual code coverage for each method was assessed, and the results are presented in Table \ref{tbl:coverage}. Our findings indicate that test programs generated by \tool~achieved significantly higher coverage on \textit{rustc}, with a relative increase of 95.92\% compared to RustSmith and 117.73\% compared to Rustlantis. Furthermore, \tool~also outperformed the traditional variable-filling mutation-based method SPE.

\begin{tcolorbox}[notitle,boxrule=0pt,colback=yellow!10,colframe=blue!20
] 
\textbf{Summary on Tools Comparison}: Compared to both generation-based and mutation-based fuzzers, \tool~outperforms in both bug discovery capability and code coverage for the Rust compiler, establishing its superiority in these aspects.
\end{tcolorbox}

\subsection{RQ3: Contribution of key components}
\label{sec:rq3}
Our methodology consists of five main components: historical code snippets triggering bugs, the \texttt{clozeMask} strategy, data augmentation, fine-tuning, and prompts based on code context. Therefore, we have designed five variants of \tool~ to assess their contributions to our approach.

\noindent $\bullet$ \textbf{\tool$_{nh}$}: It leverages regular Rust code snippets rather than the historical bug-triggering code snippets for fine-tuning the Incoder model in \tool. With the fine-tuned Incoder model, we then generate new test code using the \texttt{clozeMask} strategy by infilling the masked regular Rust code snippets. 

\noindent $\bullet$ \textbf{\tool$_{nc}$}: It adopts a random line masking strategy rather than the \texttt{clozeMask} strategy for the historical bug-triggering code snippets. In other words, a line is randomly chosen for masking before utilizing the fine-tuned Incoder model for code completion.

\noindent $\bullet$ \textbf{\tool$_{nf}$}: It utilizes an untuned Incoder model to infill the historical bug-triggering code snippets which are masked by the \texttt{clozeMask} strategy.

\noindent $\bullet$ \textbf{\tool$_{na}$}: It leverages a fine-tuned model using non-augmented data and utilizes the fine-tuned model to populate historical bug-triggering code snippets masked by the \texttt{clozeMask} strategy.

\noindent $\bullet$ \textbf{\tool$_{nm}$}: Instead of employing any mask strategy, it directly utilizes a fine-tuned Incoder model to solely generate Rust code from scratch as test cases.

In RQ3, we consider the results of \tool~as the baseline. We investigated the code coverage and bug-hunting results achieved by various implementation variants in this experiment. 
Following the experimental setup consistent with RQ2, we evaluated the coverage of various variants within a 24-hour timeframe on the stable version 1.73 of \textit{rustc}.
The code snippet dataset utilized by \tool$_{nh}$ is sourced from the Rust open-source projects, specifically the nomicon~\cite{nomicon}, rust-by-example~\cite{rust-by-example}, and rust-cookbook~\cite{rust-cookbook} repositories. It should be noted that the coverage of all the original test programs (including test cases that have historically triggered defects and the \textit{rustc} test suite) is 73.64\%. Considering that our method employs random algorithms when generating new test programs, not all test programs are utilized.

The results are presented in Table \ref{tbl:coverage}. 
It can be observed that in the absence of historical-triggered defective code, the performance of \tool$_{nh}$ significantly deteriorates compared to the baseline. This validates the effectiveness of incorporating historical-triggered defective code in our approach.
The strategy based on \texttt{clozeMask} is equally important. Although \tool$_{nc}$ relying on randomly masked fillings also utilizes the context of Rust code snippets to assist the LLM in generating test cases, it does not outperform the baseline method in terms of coverage and bug-hunting. This is because the parentheses structure better preserves the syntactic and semantic integrity of the original code. Additionally, many of Rust's complex syntax features are composed of parentheses structures, as demonstrated in Section~\ref{subsec:preStudy}.
\tool$_{nf}$ and \tool$_{na}$ exhibit a relatively minor decrease in both code coverage and bug-finding ability compared to the baseline. This could be attributed to the scarcity of code that triggers Rust compiler defects in the training corpus used to train the original LLM. As a result, the effects of fine-tuning and data augmentation components are not as evident as we had anticipated.
Another notable observation is that without the contextual guidance provided by Rust code snippets, relying solely on the generative capability of LLM, \tool's performance is significantly compromised.
The programs generated by \tool$_{nm}$ consist mostly of invalid characters and incorrect syntax, indicating the necessity of code context to guide LLM in generating test cases for complex compiler scenarios.

\begin{tcolorbox}[notitle,boxrule=0pt,colback=yellow!10,colframe=blue!20] 
\textbf{Summary on Contributions of Components}: In conclusion, our comprehensive evaluation indicates that the key components of \tool~exhibit remarkable effectiveness, particularly in enhancing code coverage and bug-hunting capabilities.
\end{tcolorbox}

\begin{table}
\setlength{\abovecaptionskip}{0cm} 
\caption{Code coverage achieved by different variants of \tool. The column labeled \textit{\#Prog.} represents the number of test cases generated in 24h, \textit{Code Cov.} represents code coverage and \textit{Change} represents the relative percentage decrease in coverage obtained by each variant compared to the coverage achieved by \tool.}
\label{tbl:coverage.component}
\renewcommand{\arraystretch}{1.25}
\setlength{\tabcolsep}{3mm}{
\begin{center}
\begin{tabular}{l|ccrc}
\hline
\rowcolor[HTML]{FFFFFF} 
                                    & \textit{\#Prog.}          & \textit{Code Cov.}   &\textit{Change} &\textit{Bugs}\\ \hline
\rowcolor[HTML]{FFFFFF} 
\cellcolor[HTML]{FFFFFF}\tool       &18775              & \textbf{65.67\%}  & - &\textbf{11} \\ 
\cellcolor[HTML]{FFFFFF}\tool$_{nh}$       &19001              & 42.34\%  & 35.52\% &0 \\ 
\rowcolor[HTML]{FFFFFF} 
\tool$_{nc}$                           &18984                          & 50.01\% &23.85\%$\downarrow$ &2\\
\rowcolor[HTML]{FFFFFF} 
\tool$_{nf}$                           &18890                          & 61.89\% &5.76\%$\downarrow$ &9\\
\rowcolor[HTML]{FFFFFF} 
\tool$_{na}$                           &18779                          & 60.46\% &7.66\%$\downarrow$ &10\\
\rowcolor[HTML]{FFFFFF} 
\tool$_{nm}$                           &22736                           & 7.01\% &89.33\%$\downarrow$ &0\\\hline
\end{tabular}
\end{center}
}
\vspace{-0.5cm}
\end{table}

%% file: content/discussion.tex
\begin{figure*}
    \centering
    \begin{subfigure}[t]{0.25\textwidth}
        \includegraphics[width=\textwidth]{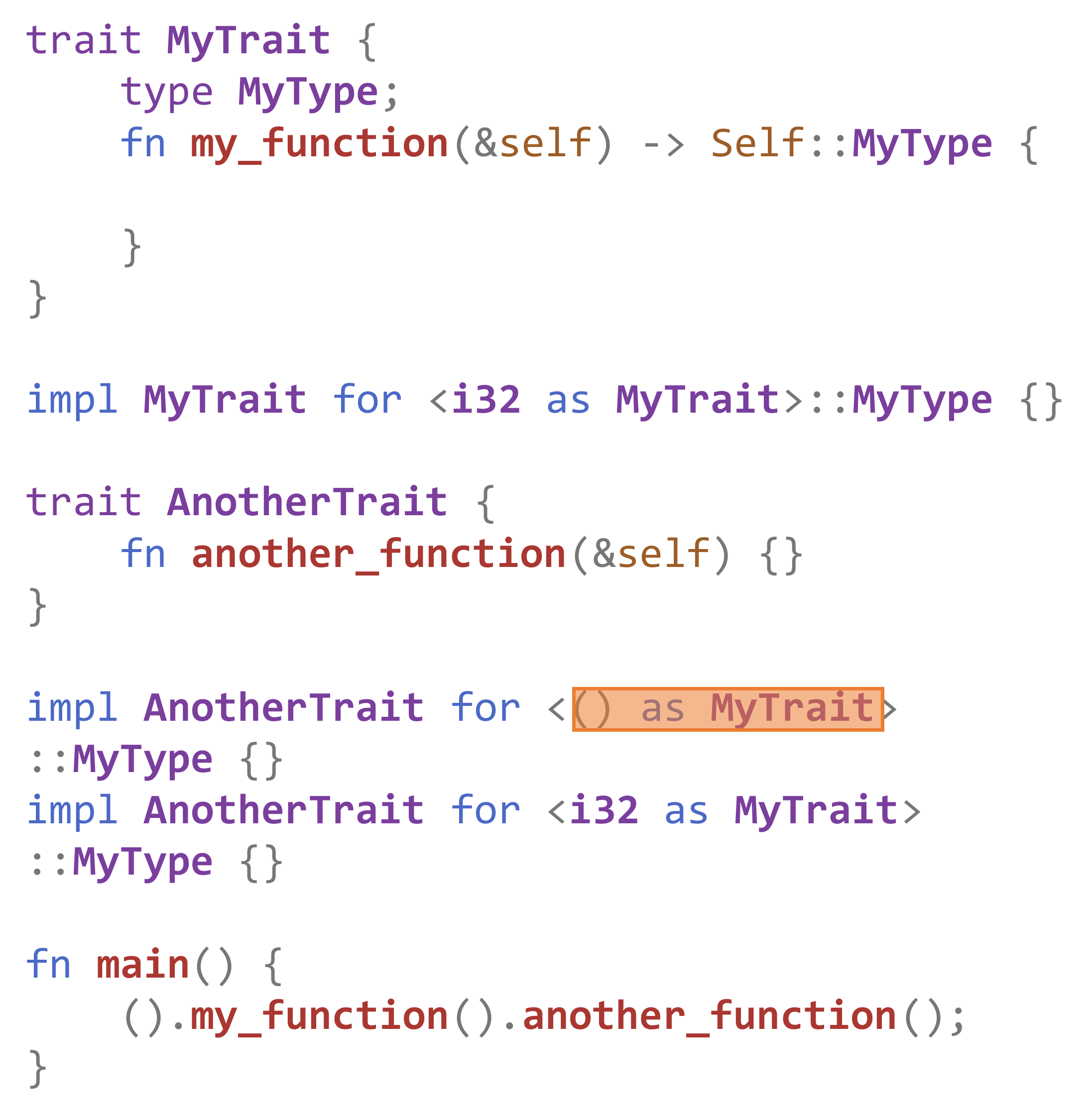}
        \caption{\href{https://github.com/rust-lang/rust/issues/116287}{\textcolor{mypurple}{\textbf{\#Issue116287}}}}
        \label{rq4.exp1}
    \end{subfigure}
    \begin{subfigure}[t]{0.20\textwidth}
        \includegraphics[width=\textwidth]{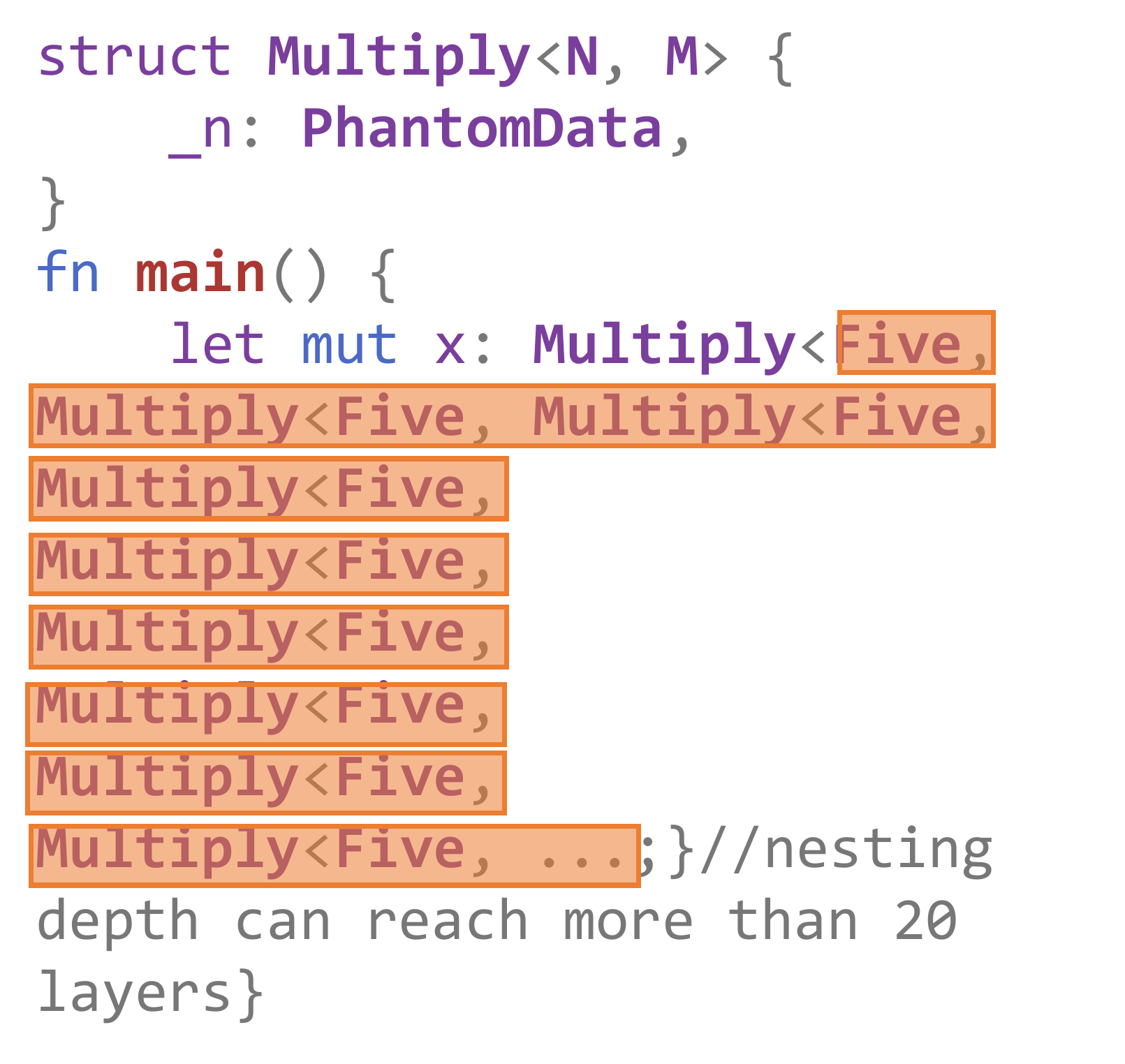}
        \caption{\href{https://github.com/rust-lang/rust/issues/116647}{\textcolor{mypurple}{\textbf{\#Issue116647}}}}
        \label{rq4.exp2}
    \end{subfigure}
    \begin{subfigure}[t]{0.25\textwidth}
        \includegraphics[width=\textwidth]{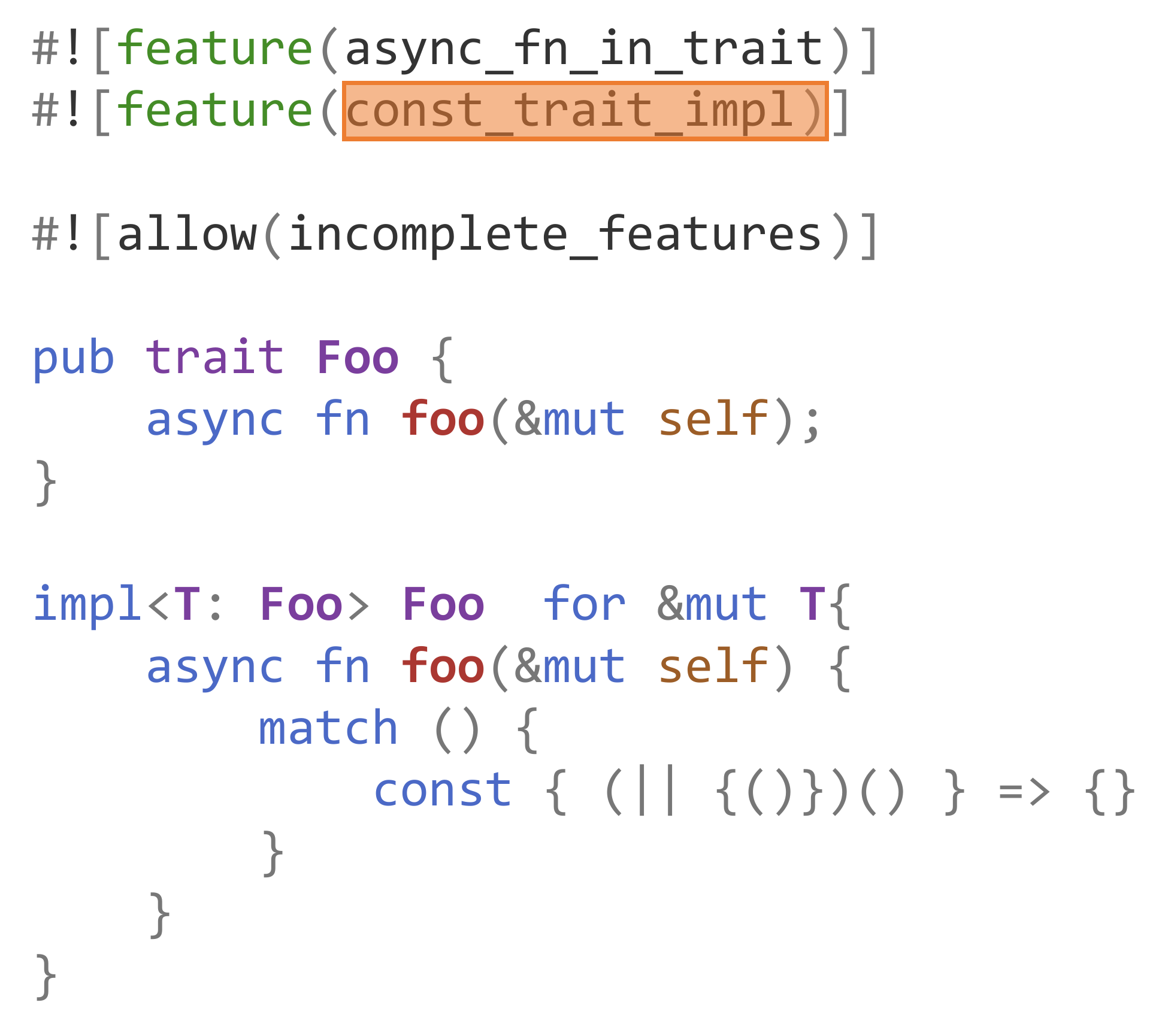}
        \caption{\href{https://github.com/rust-lang/rust/issues/117634}{\textcolor{mypurple}{\textbf{\#Issue117634}}}}
        \label{rq4.exp3}
    \end{subfigure}
    \begin{subfigure}[t]{0.26\textwidth}
        \includegraphics[width=\textwidth]{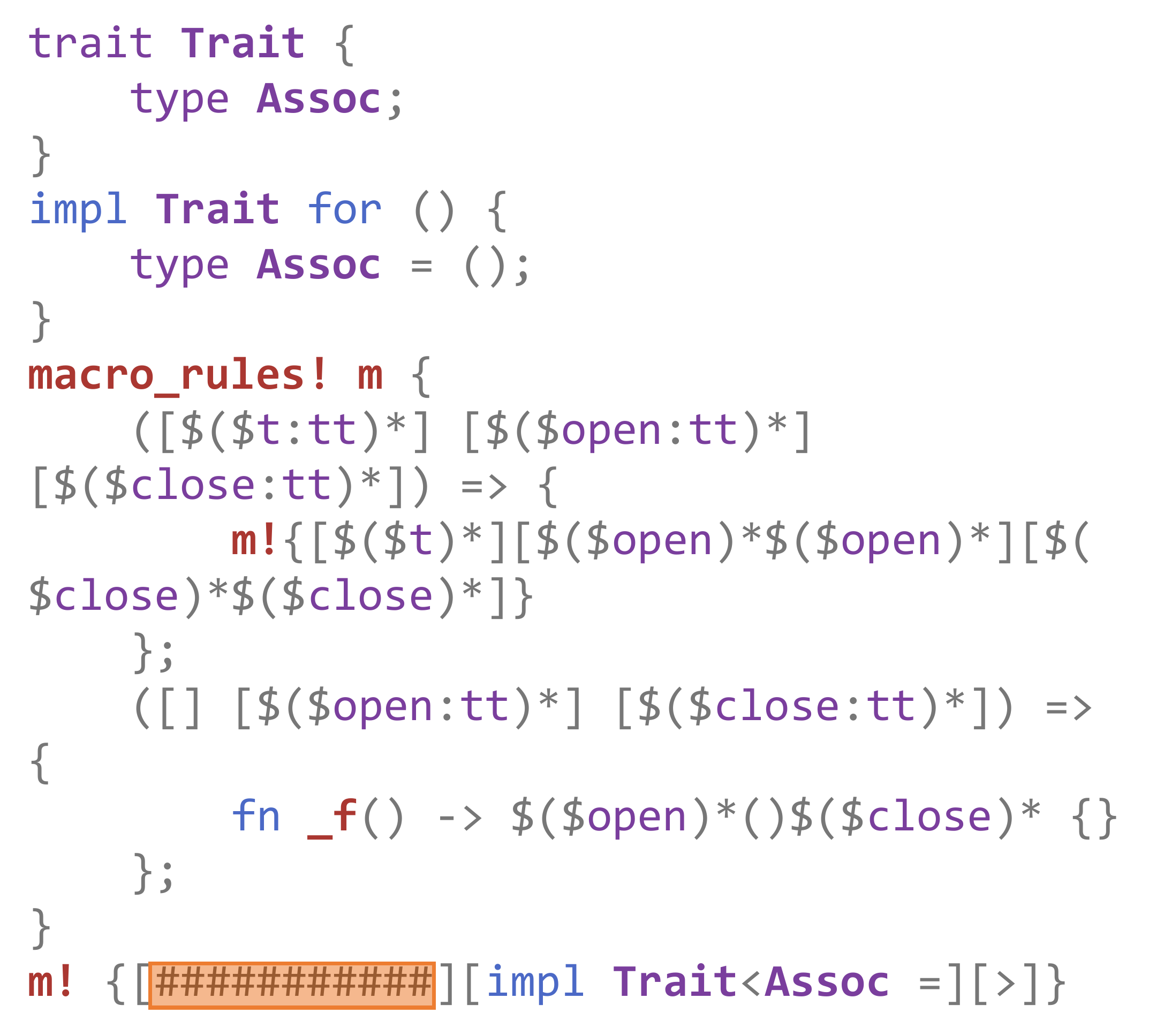}
        \caption{\href{https://github.com/rust-lang/rust/issues/116681}{\textcolor{mypurple}{\textbf{\#Issue116681}}}}
        \label{rq4.exp4}
    \end{subfigure}
   
    \caption{Example cases generated by \tool~that can reveal bugs in \textit{rustc}. The portion generated using the \texttt{clozeMask} strategy is highlighted with light orange.}
    \label{rq4.exp}
    \vspace{-0.7cm}
\end{figure*}
In this section, we mainly explore the complexity of the test cases generated by the \tool, the influence of the selected large language model on the \tool~'s performance, and the potential threats to the validity of our research findings.
\subsection{Complexity and Case Analysis of Rust Test Code Generated by \tool}
To evaluate the complexity of the code generated by the \texttt{clozeMask} strategy in \tool, we provide elucidation through two code evaluation metrics and exemplary analysis. 
\subsubsection{Metrics}
We analyzed the 18775 test cases generated by \tool~over a 24-hour period, focusing on the code snippets filled by the \texttt{clozeMask} strategy. The two code evaluation metrics used are:

\textbf{\textit{Nested Parenthesis Structure.}} Our observations indicate that the average nesting depth of the \texttt{clozeMask-generated} segments is approximately 2, with the maximum nesting depth exceeding 20 levels.

\textbf{\textit{Diversity of features.}} The \texttt{clozeMask} strategy generated 1230 different features. The most frequently occurring ones are \texttt{type\_alias\_impl\_trait} (7.07\%), followed by \texttt{const\_generics} (6.28\%), \texttt{generic\_associated\_types} (5.35\%), \texttt{generic\_const\_exprs} (4.42\%), and \texttt{const\_fn} (3.34\%). Among these, \texttt{generic\_const\_exprs} triggers the most Internal Compiler Errors (ICEs) (3/11), due to it being a compiler feature still under development.
\subsubsection{Examples of reported bugs}
Here, we present a case study to illustrate the key characteristics of the test cases generated by \tool~capable of triggering bugs.
Fig.~\ref{rq4.exp} illustrates 4 bugs discovered in the \textit{rustc} compiler by the test cases generated through \tool.
\begin{adjustwidth}{0.2cm}{0cm}
\begin{enumerate}[label=\alph*.,leftmargin=*,nosep]
    \item \textit{\textbf{Grammatically correct statements based on context}}: Fig.~\ref{rq4.exp1} is a Hang issue. The introduction of the ``\texttt{() as MyTrait}" statement has added unexpected complexity to the interaction between the ``\texttt{MyTrait}" and ``\texttt{AnotherTrait}" implementations, causing the compiler to consume excessive resources when processing this part of the code, ultimately leading to a deadlock. 
    \item \textit{\textbf{Complex nested structures}}: Fig.~\ref{rq4.exp2} shows that the \texttt{clozeMask} strategy introduces a complex nested ``\texttt{Multiply}" type structure, which overwhelms the compiler's type inference and parsing capabilities. The circular nested types generated by \texttt{clozeMask} exceed the compiler's processing limits. Consequently, the code's design surpasses the compiler's ability to handle it in a reasonable timeframe.
    \item \textit{\textbf{Different features}}: Fig.~\ref{rq4.exp3} is an ICE issue. The combination of the ``\texttt{const\_trait\_impl}" feature and the use of a constant closure within an async trait method implementation by the \texttt{clozeMask} strategy overloads the compiler's capabilities. The constant expression evaluation interacts poorly with async function handling, leading to the observed Internal Compiler Error. This complex interaction between language features has triggered an edge case in the Rust compiler. 
    \item \textit{\textbf{Uncommonly-written code snippet}}: Fig.~\ref{rq4.exp4} is a hang issue caused by the ``\texttt{m!}" macro's recursive nature. Each invocation of ``\texttt{m!}" adds more tokens to the macro's input without any termination condition. The ``\texttt{\#\#\#\#\#\#\#\#\#\#\#}" triggers repeated calls to ``\texttt{m!}", resulting in infinite recursion. The compiler cannot handle the endless macro expansion, leading to a hang.
\end{enumerate}
\end{adjustwidth}

The \texttt{clozeMask} strategy in \tool~generates complex test cases for the Rust compiler due to their deep nesting and use of diverse features. These complex cases can uncover compiler bugs by introducing unexpected interactions or edge cases. The findings suggest that \tool~can provide valuable inputs for improving the Rust compiler's robustness.
\subsection{Selection of Different LLMs}
\begin{table}[b]
\setlength{\abovecaptionskip}{0cm} 
\vspace{-0.2cm}
\caption{Evaluation of Compilation Success and Code Coverage for Various LLMs in Rust Code Infilling Tasks}
\vspace{-0.3cm}
\label{tbl:modelspass}
\begin{center}
\setlength{\tabcolsep}{5mm}
\begin{tabular}{l|cc}
\hline
          & \textit{Pass Rate }                              & \textit{Code Coverage}                                                \\ \hline
StarCoder & 2.00\%                                   & 10.8\%                                                                     \\
CodeShell & 3.00\%                                   & 15.6\%                                                                    \\
Incoder   & \cellcolor[HTML]{FFCCC9}\textbf{11.00\%} & \cellcolor[HTML]{FFCCC9}\textbf{36.1\%} \cellcolor[HTML]{FFCCC9} \\ \hline
\end{tabular}
\end{center}
\vspace{-0.4cm}
\end{table}
We propose the \tool~ framework, which can theoretically be combined with any pre-trained code generation LLM that supports the Rust language. However, our experimental results indicate that the model's performance on mainstream datasets (e.g., HumanEval~\cite{openai}) does not necessarily make it more suitable for our \tool~ framework. Specifically, we evaluated the code infilling capabilities of the state-of-the-art open-source models for Rust code infilling tasks, namely Incoder~\cite{incoder}, StarCoder~\cite{starcoder}, and CodeShell~\cite{codeshell}, on three Rust official tutorial datasets (nomicon~\cite{nomicon}, rust-by-example~\cite{rust-by-example} and rust-cookbook~\cite{rust-cookbook}).
We randomly selected 100 code snippets from these datasets, extracted their code templates (i.e., randomly masking the contents within pairs of brackets), and then used the LLM to fill in the missing code. The filled code was compiled using the \textit{rustc} (stable version 1.73). The results, as shown in Table \ref{tbl:modelspass}, indicate that the code filled by Incoder achieved the highest pass rate (with 11 programs successfully compiling and executing), while StarCoder and CodeShell had pass rates of 2\% and 3\%, respectively. Our manual analysis reveals that in the generated test programs, the latter two models incorrectly classify many Rust program errors as Python programs, resulting in the generated code being inconsistent with Rust syntax rules.

These results suggest that Incoder has greater potential compared to the other models for the task of Rust code filling. This is because Incoder employs a strategy of random masking between code snippets during the model training phase~\cite{incoder}, which enhances its ability to fill in code within the middle of existing code. Therefore, Incoder is better suited for our \tool~ framework compared to other LLMs.
We also attempted to use GPT-4o~\cite{gpt4o} to directly generate test code for the Rust compiler. Within 24 hours, 11,883 test programs were generated, with a compilation pass rate reaching 32.22\%, surpassing the lightweight open-source models listed in Table \ref{tbl:modelspass}. However, no bugs were found. This is consistent with our observation in Section \ref{rq4.exp4}.
Considering that ~\tool is a prototype framework, using Incoder can achieve significantly better results than the current Rust compiler testing tools discussed in Section~\ref{sec:evaluation}. We believe that applying a better model to our framework might yield better results in Rust compiler testing.

\subsection{Threats to validity.}
The primary threats to internal validity are bugs in implementation and experimentation, which have been mitigated through rigorous code review. The main threats to external validity come from the chosen subject systems, which have been reduced by selecting the two most popular Rust compilers, \textit{rustc} and \textit{mrustc}~\cite{Rustsmith,RustStackOverflow,CLPRust}. We have also used widely used metrics from prior fuzzing studies~\cite{10.1109/ICSE48619.2023.00018} to ensure real bug detection and code coverage.

%% file: content/related.tex
\mypara{Generative fuzzers.}
Generative fuzzers which construct test cases based on random syntax rules~\cite{10.1109/WETICE.2014.33,10.1109/CISIS.2013.99,10.1145/3062341.3062349} or templates~\cite{10.1109/WETICE.2014.33,10.1145/3092282.3092285} and determine whether a bug is detected through differential testing or compilation crash is a common approach in compiler testing. For instance, tools like Csmith~\cite{Csmith} and Yarpgen~\cite{Yarpgen} are utilized to synthesize test cases for C language compilers, while CLSmith~\cite{CLSmith} is employed for OpenCL testing. RustSmith~\cite{Rustsmith} is a prominent generative fuzzing method designed for testing the Rust compiler. It generates random Rust test code by constructing abstract syntax trees (AST) that adhere to the Rust grammar. The code generation process is context-aware, ensuring compliance with Rust’s typing rules and semantics. In addition, Dewey et al.~\cite{CLPRust} have explored the automatic generation of Rust programs by formulating problems using Constraint Logic Programming (CLP). However, their work is limited to the Rust type system module. In contrast, \tool~aims to uncover bugs in the Rust compiler by generating diverse test cases based on historical test programs that reveal bugs. These generated test cases exhibit greater diversity and cover a broader range of compiler components. To the best of our knowledge, we are the first to undertake such an endeavor on the Rust compiler.
\mypara{LLM-based fuzzers.}
With the rise of Large Language Models (LLMs), LLMs have been utilized for software testing.
WhiteFox~\cite{whitefox} explores using LLMs for assisting white-box testing of compilers, while KernalGPT~\cite{KernelGPT} attempts to leverage LLM for enhancing kernel fuzzing.
Methods such as TitanFuzz~\cite{TitanFuzz} and FuzzGPT~\cite{FuzzGpT} have started incorporating LLMs into test case generation, but primarily focus on testing Python API libraries. Recently, the work by Gu~\cite{10.1145/3611643.3617850} and Fuzz4all~\cite{UniverseFuzz} demonstrated LLM effectiveness in generating compiler test cases, but are limited to mature programming languages like Python, Java, Go, and C++. 
To our knowledge, there is no research discussing the application of LLM-based fuzzers in Rust compiler testing, likely due to limited Rust training corpora and complex syntax.
For instance, TitanFuzz designed methodology for testing Python API libraries, generating test cases from scratch using LLM and applying masking and filling transformations based on specific API or function calls. However, this approach cannot be directly applied to Rust due to ongoing language development and lack of mature corpus, making it challenging to generate test code solely using LLM without contextual information .
Furthermore, the masking and filling techniques in TitanFuzz are limited to specific Python libraries and APIs, inapplicable to Rust. Some other works explored masking and filling methods, but predominantly for APR tasks~\cite{apr}. FuzzGPT and HistFuzz~\cite{10.1109/ICSE48619.2023.00018} propose combining LLM with historically triggered code snippets to generate comprehensive test cases. However, these approaches rely on direct generation using LLM, which performs well on mature programming languages like Python and relatively SMT formula languages but is unsuitable for Rust, which is a relatively niche language, as discussed in section~\ref{sec:rq3}.
Inspired by existing methodologies, we carefully selected an LLM suitable for Rust code completion tasks. By fine-tuning the LLM with input from historically triggered code snippets and utilizing Rust's bracket structures for masking and filling, we leverage the contextual information present in the original Rust corpus. Our approach focuses on generating targeted test cases specifically for the Rust compiler.

%% file: content/conclusion.tex
 We propose and implement \tool, the first approach for fuzzing Rust compilers by leveraging historical bug-triggering inputs. Our method involves using bug-triggering code collected from past incidents to generate cloze-masked code snippets and utilizing the contextual understanding capability of the LLM to perform fill-in-the-blank tasks and generate new test cases. To evaluate the \tool's effectiveness, we conducted a hunting campaign exposing real bugs in two widely used Rust compilers, \textit{rustc} and \textit{mrustc}. \tool~successfully detected \confirmed~confirmed bugs in \textit{rustc} and \textit{mrustc}, \fixed~already fixed by the developers. Furthermore, \tool~outperformed state-of-the-art fuzzers in terms of code coverage and effectiveness. Our experimental results demonstrate potential for utilizing historical information for in Rust compiler testing.